\documentclass[twocolumn,showpacs,preprintnumbers,aps,prl,amsmath,amssymb]{revtex4}

\usepackage{graphicx}
\usepackage{dcolumn}
\usepackage{bm}

\begin{document}

\title{Local Blockade of Rydberg Excitation in an Ultracold Gas }

\author{D. Tong}
\author{S.M. Farooqi}
\author{J. Stanojevic}
\author{S. Krishnan}
\author{Y.P. Zhang}
\author{R. C\^ot\'e}
\author{E.E. Eyler}
\author{P.L. Gould}
\affiliation{Physics Department, University of Connecticut,
Storrs, CT 06269}
\date{\today}

\begin{abstract}
In the laser excitation of ultracold atoms to Rydberg states, we
observe a dramatic suppression caused by van der Waals
interactions. This behavior is interpreted as a local excitation
blockade: Rydberg atoms strongly inhibit excitation of their
neighbors. We measure suppression, relative to isolated atom
excitation, by up to a factor of 6.4. The dependence of this
suppression on both laser irradiance and atomic density are in
good agreement with a mean-field model. These results are an
important step towards using ultracold Rydberg atoms in quantum
information processing. \end{abstract}

\maketitle
%
%
The possibility of a computer that operates according to the
principles of quantum mechanics has attracted growing interest
from a variety of research fields \cite{Macchiavello, Nielsen}.  A
number of possible implementations are being investigated,
including solid-state systems, nuclear magnetic resonance, cavity
quantum electrodynamics, trapped dipolar molecules, trapped ions,
and trapped neutral atoms. A key element to any successful system
is the ability to control the coherent interactions between the
fundamental building blocks (qubits). Highly-excited Rydberg atoms
with principal quantum numbers $n \gtrsim 30$ have the advantage
that they interact quite strongly with each other, allowing
information to be exchanged quickly \cite{Jaksch}. Here we report
an important advance towards using ultracold Rydberg atoms in
quantum computing. We observe that the laser excitation of a
macroscopic sample of ultracold atoms to high-lying Rydberg states
can be dramatically suppressed by their strong long-range
interactions. This leads to a local blockade effect, where the
excitation of one atom prevents excitation of its neighbors. Our
observations agree well with a model based on mean-field
interactions.

In a high-$n$ Rydberg state, the electron spends most of its time
quite far from the nucleus \cite{Gallagher}.   As a result, the
energy of this highly-excited state is very sensitive to external
perturbations, including those caused by neighboring Rydberg
atoms. A system of two ultracold Rydberg atoms, subject to these
long-range interactions, has been proposed as a possible
realization of a quantum logic gate \cite{Jaksch, Protsenko}.
Rydberg states combine the advantages of long radiative lifetimes
and strong long-range interactions, allowing information to be
exchanged before decoherence sets in, even when the atoms are
sufficiently separated to allow individual addressing. If the
atoms are ultracold, they can be highly localized, e.g., in the
sites of an optical lattice, allowing control of their
interactions and efficient detection of their quantum state. An
outstanding challenge is the assurance that at most a single
Rydberg atom is produced at a given site. Towards this end, the
concept of an excitation blockade has been proposed \cite{Lukin,
Saffman}.  With multiple atoms occupying a sufficiently localized
site, the strong Rydberg-Rydberg interactions allow at most one
Rydberg excitation. Further excitations are blocked by the large
energy level shifts that push the resonant frequencies outside the
laser bandwidth.

For two atoms in $n$p states, separated by a distance $R$, the
$C_6/R^6$ van der Waals (vdW) interaction dominates at long range
\cite{Boisseau, Marinescu}. The rapid $n^{11}$ scaling of the
$C_6$ coefficient indicates the advantage of using high-$n$
Rydberg states. The original proposals for quantum gates\cite
{Jaksch} and an excitation blockade \cite{Lukin} with Rydberg
atoms involved finite electric fields which mixed the angular
momentum states, giving rise to a $C_3/R^3$ dipole-dipole
interaction. The vdW interaction, which we have employed, has the
important advantage of providing nonzero energy level shifts for
every molecular state regardless of the orientation of the atoms.
By contrast, dipole interactions vanish for certain orientations,
reducing the efficiency of the dipole blockade mechanism.

\begin{figure}
 \includegraphics[width=\linewidth]{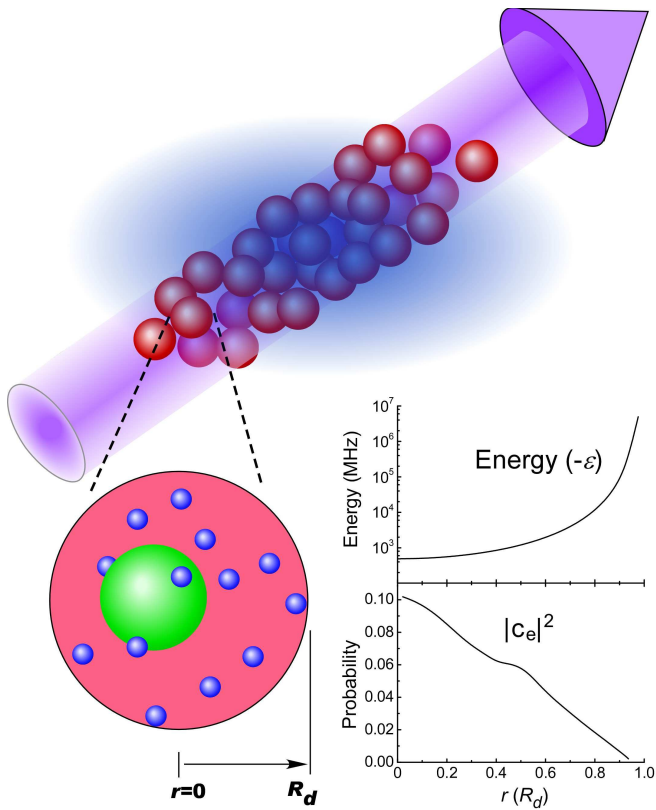}
 \caption{\protect\label{overview} Schematic of the experiment.
 Top: a UV laser beam illuminates a sample of ultracold
 $^{85}$Rb atoms in order to excite Rydberg atoms.  Our
 theoretical model divides the region into domains of radius
 $R_d$. Zoom: Within a given domain, any one of the numerous
 ground-state atoms may be excited into a Rydberg state.
Inset: Because of strong Rydberg-Rydberg vdW
 interactions, the mean-field energy shift $\varepsilon$ depends
 on the particular location, as does the excitation probability
 $|c_e|^2$.
 Atoms near the domain center are less shifted and their
 excitation is more probable than those near the periphery. The
 graphs correspond to $n$=80 atoms at $\rho = 6.5 \times 10^{10}$
 cm$^{-3}$ and scaled irradiance $I$ = 0.187 MW/cm$^2$.}
\end{figure}

The basic idea of the experiment is shown in Fig. \ref{overview}.
We start with a sample of about $10^7$ ultracold $^{85}$Rb atoms
in the 5$s$ ground state and illuminate them with a UV pulse
resonant with a transition to an \textit{np} Rydberg state
\cite{Farooqi}. Once a Rydberg atom is produced, the strong vdW
interaction shifts the energy levels of neighboring atoms,
suppressing their excitation. As a result, each Rydberg atom
resides in a domain, within which no other Rydberg excitation is
possible. The higher the atomic density, the more atoms will be in
each domain, and the larger will be the factor by which the
overall excitation fraction of the macroscopic sample is
suppressed. It is this suppression that we observe in the
experiment.

The ultracold sample of $^{85}$Rb atoms, with a peak density up to
$10^{11}$ cm$^{-3}$ and a temperature of $\sim$100 $\mu$K, is
provided by a diode-laser-based vapor-cell magneto-optical trap
(MOT). Trapped atoms in a specific hyperfine level of the ground
state are excited to $np_{3/2}$ states with $n$=30-80 by a
narrowband 297 nm laser pulse of 8.6 ns duration (FWHM), generated
by frequency doubling of a pulse-amplified cw laser as in our
earlier work \cite{Farooqi}. The bandwidth of about 100 MHz,
measured by scanning over the 30$p$ resonance, is about twice the
Fourier transform limit due to frequency chirping in the pulsed
amplifier \cite{Melikechi}. In order to excite the highest density
region, the UV light is focused into the MOT cloud, yielding a
cylindrical excitation volume $\sim$500 $\mu$m long and $\sim$220
$\mu$m in diameter (FWHM). Within 60 ns after the laser pulse, a
$\sim$1500 V/cm electric field is applied, ionizing the Rydberg
atoms and accelerating the ions towards a microchannel plate (MCP)
detector. The trapping light is turned off when the UV pulse
arrives in order to prevent direct photoionization from the 5$p$
level.

The MCP is calibrated using two methods.  The first is based on
the signal from near-threshold photoionization of the 5$s$ ground
state. The measured density distribution in the MOT, obtained from
the trapped-atom fluorescence profile, is combined with the
measured UV beam parameters and the known photoionization cross
section\cite{Marr, Ciampini} to calculate the number of photoions
per laser pulse. The second method is based on the $n$=30 Rydberg
signal, which behaves as isolated-atom excitation at all
intensities used, because for $n$=30 the vdW interaction is
relatively weak. A linear fit is combined with calculations
utilizing the $5s\rightarrow 30p$ oscillator strength of Ref.
\cite{Shabanova}, including the effects of a linear laser
frequency chirp corresponding to the observed bandwidth. The two
calibrations agree within 2\%, confirming that we can make
accurate quantitative predictions of excitation probabilities in
the absence of a blockade.

\begin{figure}
\includegraphics[width=\linewidth]{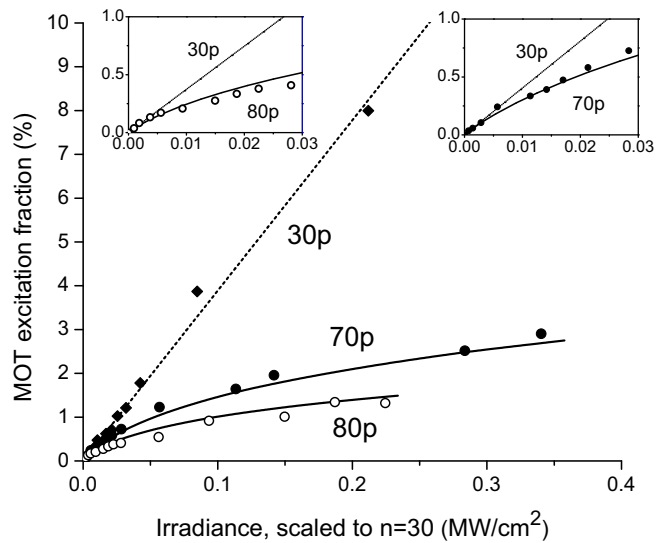}
\caption{\protect\label{ndep}  Comparison of Rydberg excitation
for the unblockaded (isolated atom) 30$p$ state and the blockaded
70$p$ and 80$p$ states at a peak density of $6.5 \times 10^{10}$
cm$^{-3}$. Irradiances are scaled by ($n$*/30*)$^3$ to account for
the $n$ dependence of the electric dipole transition probability.
Insets show the region near the origin with an expanded scale. The
dashed line for $n$=30 is a least-squares fit to the data, while
the solid curves for $n$=70 and $n$=80 are theoretical
predictions, using a single adjustable scaling parameter $\alpha$
as described in the text. To account for the calculated effects of
slight variations in the MOT cloud parameters, the $n$=30 results
are multiplied by 1.29 ($n$=70 inset), 1.19 ($n$=80 inset) and
1.24 (main figure, using the average correction for $n$=70,80).}
\end{figure}

The dependence of the Rydberg signal on the peak UV irradiance is
shown in Fig. \ref{ndep} for $n$=30, 70 and 80. The signal plotted
is the fraction of the entire MOT sample that is excited. For each
$n$, the irradiance values are scaled to $n$=30 by the factor
(30*/$n$*)$^3$ in order to account for the decrease in transition
strength with increasing $n$. Here $n$*=$n-\delta$, and
$\delta=2.6415$ is the quantum defect for $p_{3/2}$ states
\cite{Lorenzen}. Note that the $n$=30 saturation intensity for
isolated atoms, defined as that required for an unchirped
$\pi$-pulse in the center of the beam, is 0.36 MW/cm$^2$. With
this irradiance scaling, the various $n$'s would fall on a
universal isolated atom excitation curve if the Rydberg levels
were unshifted by atomic interactions. This is seen to be the case
for the very lowest intensities, at which the Rydberg atoms are
sufficiently sparse that interactions between them are negligible.

The salient feature of Fig. 2 is the dramatic suppression of
Rydberg excitation for $n$=70 and 80 relative to the isolated atom
($n$=30) excitation curve. As expected, the suppression is larger
for $n$=80 due to its stronger vdW interaction, reaching a factor
of 6.4 at the highest intensities shown.

    We model the suppression of Rydberg excitation by solving the
Bloch equations for the ground-state and excited-state amplitudes,
$c_g$ and $c_e$, of a given atom. The key point is to include the
energy level shift $\varepsilon$ due to interactions with nearby
Rydberg atoms. If we consider an $np_{3/2}$ Rydberg atom located
at $\bm{r}_i$, labelled $\left| {p_i } \right\rangle $, the
first-order shift due to its interaction $\hat V_{\rm int}$ with
$\left| {p_k } \right\rangle $ is $\epsilon_{ik} = \left\langle
{p_i p_k } \right|\hat V_{\rm int} ({\bf{r}}_i  - {\bf{r}}_k
)\left| {p_i p_k } \right\rangle $. At large separations, this
shift is dominated by the vdW interaction corresponding to a pair
of molecular states $ ^1\Sigma _g^+$ and $^3\Sigma _u^+$, labeled
below as $\lambda$ = 1 and 2 respectively \cite{Boisseau,
Marinescu, Farooqi}. For $n$=70, $C_6 = 2.64 \times 10^{22}$ a.u.
for both states.

    For simplicity, we consider a spherical domain of radius $R_d$
and volume $V_d$ which contains several atoms, but by definition,
only a single Rydberg atom $\left| {p_i } \right\rangle$. All
other excited atoms, outside the domain, contribute to the total
shift of this Rydberg level $\varepsilon_i  = \Sigma_{k \ne i}
\,\varepsilon_{ki}$. The shift of $\left| {p_i } \right\rangle $,
and therefore its probability of excitation, depends upon its
location within the domain, as shown in the insets to Fig. 1. At
the center of the sphere, the distance from any other excited atom
is maximized, and the shift is therefore minimized. At the
periphery of the domain, the proximity of external Rydberg atoms
increases the shift, leading to a stronger suppression of
excitation. The domain radius $R_d$ is determined from the
condition
\begin{equation}
\rho \int _{V_d } d^3 {\bf{r}}|c_e ({\bf{r}},t)|^2  = 1,
\label{Eq1}
\end{equation}
 where  $\rho$, the local density of atoms, is
assumed uniform within the domain. The amplitude $c_e({\bf{r}},t)$
of an atom located at $\bf{r}$ depends on its shift $\varepsilon
({\bf{r}},t)$. In our mean-field model, we calculate this shift by
replacing the discrete sum by an integral over the excited atoms
outside the domain ($V'  = V-V_d)$ and considering their density
$\rho_e(t)$ to be locally uniform:
\begin{equation}
\varepsilon ({\bf{r}},t) = \rho _e (t)\int_{V'} {d^3 r'\frac{{ -
C_6 }}{{\left| {{\bf{r}} - {\bf{r'}}} \right|^6
}}\sum\limits_{\lambda  = 1}^2 {\left| {\left\langle {{p_r p_{r'}
}}
 \mathrel{\left | {\vphantom {{p_r p_{r'} } \lambda }}
 \right. \kern-\nulldelimiterspace}
 {\lambda } \right\rangle } \right|^2 } .}
\label{Eq_eps}
\end{equation}

The radius $R_d$ and density $\rho_e$ are related
self-consistently via $\rho_e V_d = 1$. The shift of an atom
located at ${\bf{r}} = {\bf{y}}\,R_d $ can be rewritten as
\begin{equation}
\varepsilon ({\bf{y}},t) = -\tilde C_6 \,g({\bf{y}})/R_d ^6 (t),
\end{equation}
where the effective vdW coefficient, $\tilde C_6$, and the spatial
variation of the shift, $g(\bf{y})$ (with $g(0)=1$), are obtained
by numerical integration of Eq. \ref{Eq_eps}. Substituting
\begin{equation}
R_d^{-3} = \rho \int _{\left| {\bf{y}} \right| \le 1} d^3 y|c_e
({\bf{y}},t)|^2
\end{equation}
into the expression for $\varepsilon ({\bf{y}},t)$ leads to
non-linear Bloch-like equations for the time-dependent amplitudes
$c_g({\bf{y}},t)$ and $c_e({\bf{y}},t)$,

\begin{subequations}
\begin{equation}
i\frac{d}{{dt}}c_g  = \frac{\Omega}{2}e^{i\beta t^2 } c_e,
\end{equation}
\begin{equation}
i\frac{d}{{dt}}c_e  = -\rho ^2 \tilde C_6 g\left| {\int_{\left|
{\bf{y}} \right| \le 1} {d^3 {\bf{y}}|c_e |^2}} \right|^2 c_e  +
\frac{\Omega}{2}e^{ - i\beta t^2} c_g.
\end{equation}
\label{BlochLike}
\end{subequations}
Here $\beta $ characterizes the chirp of the laser pulse and
$\Omega (t)$ is the Rabi frequency.

Using the adapted zeroth order wavefunctions for the molecular
states $\lambda$ = 1 and 2 \cite{Marinescu}, and averaging over
projections $m_j$ of the excited states $\left| {n\,p_{j = 3/2}
\,m_j } \right\rangle $ for atoms outside the domain and
integrating over all possible angles of molecular axes, we find $
\tilde C_6$ = 7/60 $C_6 $. By solving numerically Eqs.
\ref{BlochLike}, we find the local density of excited atoms
$\rho_e(\rho,\Omega_0)$ after the Gaussian laser pulse of peak
Rabi frequency $\Omega_0$ has ended. To compare with the
experimental measurements, this function is averaged over the
density and laser intensity profiles in the trapped sample to
provide the overall fraction of atoms which is excited.
Uncertainties in both $\rho$ and $\tilde C_6$ are taken into
account by using the scaling factor defined via $\rho_\alpha
{\tilde C_{6,\alpha}} ^{\,\,1/2}  = \alpha \rho \tilde C_6^{\,1/2}
$.

 The data and the predictions of the model described above are compared
in Fig. 2 for both $n$=70 and $n$=80. We limit the scaled
irradiance to less than 0.35 MW/cm$^2$ because the validity of the
model is questionable when the isolated atom excitation approaches
saturation. For the curves shown, we have used a scaling factor
$\alpha$=2.4. This reflects a possible underestimate of the
absolute atomic density $\rho$ (known only to within a factor of
two) and/or the effective $C_6$ coefficient. With this scaling
factor, the agreement between the model and the data is quite
good.  For $n$=80, we observe a maximum overall suppression by a
factor of 6.4 relative to isolated atom excitation, and the theory
indicates the suppression reaches a factor of 19 in the center of
the MOT.

We have also measured the excitation fraction as a function of
atomic density at a fixed laser irradiance. The density is varied
by transiently transferring the population between the two
hyperfine levels of the ground state, $F$=2 and $F$=3, which are
separated by 3.0 GHz. Since our UV laser bandwidth is very narrow,
it excites selectively from only one of these levels. Therefore,
changing their relative population immediately preceding the UV
pulse, while keeping other parameters of the MOT fixed, varies the
effective atomic density available for excitation. This population
transfer is due to the slow ($\sim$100 $\mu$s ) optical pumping
into $F$=2 which occurs when the repumping laser is turned off
before the trapping laser. The extent of the transfer is
controlled by the time interval between switching off these two
lasers. The effective density is measured by comparing the
isolated atom signals from $F$=2 and $F$=3 to the $n$=30 state.

\begin{figure}
\includegraphics[width=\linewidth]{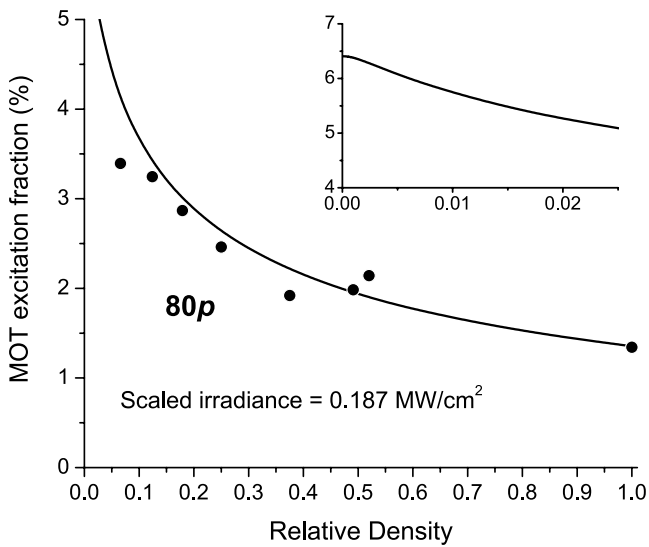}
\caption{\protect\label{densitydep} Density dependence of the
$n$=80 Rydberg signal.   Relative densities are accurate to
5-10\%, but the absolute peak density, nominally $6.5 \times
10^{10}$ cm$^{-3}$ at a relative density of 1.0, is uncertain by
as much as a factor of two.  MOT fractions are expressed in terms
of the available ground-state population in the selected hyperfine
level (see text).  The solid curve shows the mean-field
theoretical prediction, using the same scaling parameter as for
Fig. \ref{ndep}.}
\end{figure}

The measured excitation fraction for $n$=80 as a function of
atomic density is shown in Fig. \ref{densitydep} along with the
prediction of the model. Note that the excitation fraction is with
respect to the number of atoms in the appropriate hyperfine level.
The scaling factor in the model is again chosen to be
$\alpha$=2.4. Just as for the irradiance dependence in Fig.
\ref{ndep}, the data and the model agree quite well.  Note that in
the absence of a blockade, one would expect a constant excitation
fraction (here 6.4\%) instead of the rapid decrease observed.

We have verified that the suppression of excitation we have
observed is not due to Stark shifts from charges created during
the UV laser pulse. At sufficiently high intensities, we do
observe significant numbers of free electrons within 30 ns of a
resonant excitation pulse. However, at the maximum irradiance used
in the present work, 5.1 MW/cm$^2$ for $n$=80, we observe less
than 100 such electrons. In the worst case, if these electrons
were all produced during the laser pulse, the resulting average
microfield would be less than 20 mV/cm.  If all were to leave the
excitation volume, the space charge field at its edge would be
$\sim$50 mV/cm. The resulting Stark shifts for 80p are 0.9 MHz and
5.6 MHz, respectively, negligible compared to the laser bandwidth
of $\sim$100 MHz.

In conclusion, we have observed a local blockade of Rydberg state
excitation in a macroscopic sample due to strong vdW interactions.
The dependence of this dramatic suppression on laser irradiance,
atomic density, and principal quantum number are in good agreement
with a model based on a mean-field treatment of the atomic
interactions. We have observed a sample-averaged suppression by up
to a factor of 6.4 relative to isolated atom excitation.
Increasing the atomic density should result in significantly
larger suppressions. The prospects of achieving a complete
blockade in a microscopic sample, such as at a single site in an
optical lattice, look very promising. For example, with a density
of $10^{12}$ cm$^{-3}$ and a laser bandwidth of 1 MHz, we
conservatively estimate that in a spherical sample of 50 atoms,
the probability of exciting more than one atom to $n$=80 is less
than 1\%. This would represent another important step toward
quantum computing with ultracold Rydberg atoms.

This work was supported by the University of Connecticut Research
Foundation, the Research Corporation, and grants PHY-998776 and
ITR-0082913 from the National Science Foundation.

\end{document}